\documentclass[prb,twocolumn,showpacs,superscriptaddress]{revtex4}

\usepackage{amsmath, amssymb}
\usepackage{graphicx}

\newcommand{\void}[1]{}

\newcommand{\Tr}{\mathop{\mathrm{Tr}}\nolimits}
\begin{document}
\title{Phase readout of a charge qubit capacitively coupled to an open double quantum dot}

\author{Christoph Kreisbeck}
\affiliation{Institut f\"ur Theoretische Physik, Universit\"at
	Regensburg, D-93040 Regensburg, Germany}
\affiliation{Institut f\"ur Physik, Universit\"at  Augsburg,
	Universit\"atsstra{\ss}e~1, D-86135 Augsburg, Germany}
\author{Franz J. Kaiser}
\affiliation{Institut f\"ur Physik, Universit\"at  Augsburg,
	Universit\"atsstra{\ss}e~1, D-86135 Augsburg, Germany}
\author{Sigmund Kohler}
\affiliation{Instituto de Ciencia de Materiales de Madrid, CSIC,
	Cantoblanco, E-28049 Madrid, Spain}

\date{\today}


\begin{abstract}
We study the dynamics of a charge qubit that is capacitively coupled
to an open double quantum dot.  Depending on the qubit state, the
transport through the open quantum dot may be resonant or off-resonant,
such that the qubit affects the current through the open double dot.
We relate the initial qubit state to the magnitude of an emerging
transient current peak.  The relation between these quantities enables
the readout of not only the charge but also the phase of the qubit
with sufficient resolution.
\end{abstract}

\maketitle

\section{Introduction}

Quantum algorithms usually terminate with qubit readout,
i.e., a measurement of the quantum register's state.
Generally, the laws of quantum mechanics inhibit one to directly and fully
determine the wave function of the qubit from a single measurement.
With repeated projective measurements in the same basis, it is only
possible to sample the probability that the qubit is in the one or the
other of two orthogonal states.  Such destructive measurements can
nevertheless be used to demonstrate coherent oscillations by repeating
the experiment many times.  Such experiments have been performed for
superconducting qubits,\cite{Nakamura1999a,Vion2002a,Chiorescu2003a}
as well as for charge qubits implemented with double
quantum dots.\cite{Hayashi2003a, Fujisawa2004a, Gorman2005a}
Recently, the coupling of two charge qubits of the latter type
has been demonstrated,\cite{Petersson2009a} which represents a
first step toward performing gate operations.

In order to distinguish in an experiment between different charge
states of a {single quantum dot}, one may couple the dot
capacitively to a quantum point contact which acts as charge meter.
Then the current through the meter depends on the number of electrons
in the quantum dot.  This also allows one to monitor the transport of
individual electrons \cite{Gustavsson2006a} and to eventually
determine the associated full counting statistics.\cite{Fricke2007a}
It has also been proposed to couple a charge qubit to a point contact,
such that the current through the latter depends on the location of the
electron in the double quantum dot, i.e., on the state of a
charge qubit.\cite{Goan2001a, Ashhab2009b}
It is also possible to employ a voltage-biased open quantum dot as
charge meter if a nearby additional charge shifts one energy
level of the quantum dot across the Fermi surface of an attached lead,
the current depends as well on the presence of the
charge.\cite{Wiseman2001a}  When measuring the state of a charge qubit
in that way, the measurement acts back on the coherence of the qubit
which, thus, experiences decoherence and dissipation.  This means that
the qubit evolves into an incoherent mixture.  The associated
transient current allows one to infer on the initial charge state of
the qubit.

\begin{figure}[bt]
\begin{center}
\includegraphics[width=.95\columnwidth]{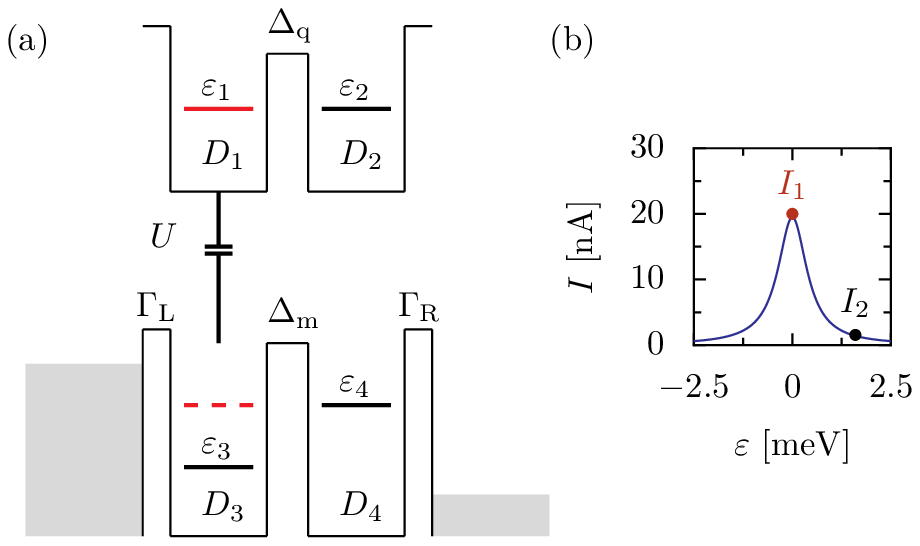}
\end{center}
\caption{\label{Ausleseidee}
(Color online)
(a) Sketch of the qubit-meter setup consisting of an open and a closed
double quantum dot.  The closed double dot ($D_1$ and $D_2$) is
occupied with one electron and forms a charge qubit.  An electron in
dot $D_1$ effectively shifts the onsite energy of dot $D_3$, such that
dots $D_3$ and $D_4$ are tuned into resonance.  Thus the current
through the open double dot ($D_3$ and $D_4$) is sensitive to the
state of the qubit.
(b) Current as a function of the meter energy bias $\varepsilon =
\varepsilon_4-\varepsilon_3$ without coupling to the qubit, $U=0$, for
inter-dot tunneling $\Delta_\mathrm{m} = 0.2\,\mathrm{meV}$ and
wire-lead coupling $\Gamma = 0.2\,\mathrm{meV}$.  The marked values
correspond to the energy shift induced by a qubit electron localized in
dot $D_1$ and $D_2$, respectively, for the interaction strength
$U=1\,\mathrm{meV}$.}
\end{figure}%
A {double quantum dot} can be used as charge meter as
well.\cite{Geszti2006a, Gilad2006a, Jiao2007a}  There the main idea is
that the monitored charge acts as gate voltage on one quantum dot,
such that the energy levels of the double dot are tuned into
resonance.  The consequence is that the conductance of the double dot
and, thus, the current increases, see Fig.~\ref{Ausleseidee}(b).  For
the parameters of this figure, the current changes by roughly a factor
10 upon qubit tunneling from dot $D_2$ to dot $D_1$.  Thus the
achievable signal-to-noise ratio for this charge measurement is
higher than the one for the single-dot meter.\cite{Gilad2006a,
Jiao2007a} Moreover, different qubit states lead to significantly
different currents, such that the measurement basis is not
fluctuating.\cite{Ashhab2009b,Ashhab2009c}
The central aim of this work is to demonstrate that the double-dot
charge meter can even be used to distinguish between qubit states with
identical population but different phase. This means that one can read
out either the charge or the phase of the qubit, depending on the
choice of the experimenter.  The phase readout proposed below
corresponds to the measurement of an observable that does not commute
with the system-meter coupling
Hamiltonian.\cite{Reuther2009a,Ashhab2009a}

\section{Qubit coupled to meter}

The setup sketched in Fig.~\ref{Ausleseidee} is described by the
Hamiltonian
\begin{align}
H = H_\text{q}+{H}_\text{m}+{H}_\text{q-m}
   +{H}_\text{m-l} +{H}_\text{leads} ,
\end{align}
where the first two terms describe the closed and the open double
quantum dot forming the qubit and the meter, respectively.  They read
\begin{align}
\label{Hq}
{H}_\mathrm{q}
={}& 
     \Delta_\mathrm{q} (c_1^{\dagger} c_2 + c_2^\dagger c_1 ) 
     = \Delta_\mathrm{q} \sigma_x
     ,
\\
\label{Hm}
{H}_{\rm m}
={}& \varepsilon_3 n_3 + \varepsilon_4 n_4
   + \Delta_\mathrm{m} (c_3^{\dagger} c_4 + c_4^\dagger c_3 ) ,
\end{align}
i.e., each dot $D_i$ is described by a single level with onsite energy
$\varepsilon_i$ and the usual fermionic creation and annihilation
operators of an electron, $c_i^\dagger$ and $c_i$, with the
corresponding electron number $n_i = c_i^\dagger c_i$.  For the
present purpose, it is sufficient to consider spinless electrons, such
that each dot can be occupied by at most one electron.
$\Delta_\mathrm{q}$ and $\Delta_\mathrm{m}$ are the inter-dot tunnel
matrix elements of the closed and the open double dot, respectively.
Since the closed double dot is occupied with a single electron, we can
write the Hamiltonian~\eqref{Hq} in pseudo-spin notation with the
basis states $|D_{1,2}\rangle = c_{1,2}^\dagger|0\rangle_\mathrm{q}$
and the usual Pauli matrices $\sigma_{x,y,z}$.  We restrict ourselves
to a qubit Hamiltonian without bias term $\propto \sigma_z$ since such
term can always be removed by proper gate voltages.

Electrons residing in dots $D_1$ and $D_3$ interact capacitively with
strength $U$ according to
\begin{equation}\label{WireQubit}
{H}_\text{q-m}= U n_1 n_3 .
\end{equation}
Notice that this qubit-meter coupling does not commute with the qubit
Hamiltonian \eqref{Hq}.  Thus the measurement is destructive and, in
particular, not of quantum non-demolition type, such that the analysis
proposed in Ref.~\onlinecite{Gambetta2007a} is not applicable here.

The meter is in addition coupled to two metallic leads which we model
as free electron gases with the Hamiltonian
\begin{equation}
H_\text{leads}
= \sum_{\ell=\rm L,R}\sum_{k}\hbar\omega_{\ell k}
  \,c^\dagger_{\ell k} c_{\ell k} ,
\end{equation}
where the operator $c^\dagger_{\ell k}$ creates an electron in state $k$ of
lead $\ell=\rm L,R$ with energy $\hbar\omega_{\ell k}$.  Henceforth, we
consider the limit of large bias voltage such that initially all
relevant states of the left lead are occupied, while those of the
right lead are empty.  Then the electron transport becomes unidirectional.
Electron tunnelling between the leads and the open double dot is
described by the Hamiltonian
\begin{equation}
{H}_\text{m-l}
=\sum_k (V_{\mathrm{L}k} c_{\mathrm{L}k}^\dagger c_3
       + V_{\mathrm{R}k} c_{\mathrm{R}k}^\dagger c_4 ) +\text{H.c.}
\end{equation}
The coupling matrix elements $V_{\ell k}$ can be subsumed
in the effective tunnel rates $\Gamma_\ell = (2\pi/\hbar)\sum_{\ell, k}
|V_{\ell k}|^2 \delta(\epsilon-\epsilon_{\ell k})$, which within a
wide-band limit are assumed to be energy independent.

In order to compute the time evolution of the system-meter setup,
we derive a master equation for the reduced density operator $\rho$ of
both double quantum dots by eliminating the leads within second-order
perturbation theory.  Starting from the Liouville-von Neumann equation
$\dot R(t)=-(i/\hbar) [H,R(t)]$ for the total density operator $R$,
we follow Ref.~\onlinecite{Kaiser2007a} and obtain by tracing out the leads
the master equation,
\begin{equation}
\label{mast}
\begin{split}
\dot\rho(t)
=& -\frac{\rm i}{\hbar}\left[{H}_{\rm s},\rho(t)\right]
 -\frac{1}{\hbar^2}\int^{\infty}_0 \!\! d\tau \Tr_\text{leads}
\\& \times
    \big\{\big[
    {H}_\text{m-l},
    \big[\tilde{H}_\text{m-l}(-\tau),
    \rho(t) \otimes\rho_\mathrm{leads}\big]\big]\big\} ,
\end{split}
\end{equation}
with $\rho_\mathrm{leads}$ being the density operator of both leads,
each in a canonical state but with different Fermi energy.
The factorization assumption for the density operator allows
evaluating the trace over the lead states such that a closed equation
for the reduced density operator of the quantum dots remains.
The first term on the right-hand side refers to the coherent dynamics
of the electrons in the two coupled double dots, while the second term
describes incoherent tunnelling between the leads and the open double
quantum dot.  The tilde denotes the interaction picture with respect
to the system Hamiltonian $H_\mathrm{s} = H_\mathrm{q}+H_\mathrm{m}
+H_\text{q-m}$. Defining the current operator in a symmetric manner,
$\mathcal{J} = (e/2)(\dot N_\mathrm{L} - \dot N_\mathrm{R})$, we
obtain the ensemble-averaged expectation value $I(t) =
(e/2)\Tr\{(\mathcal{J}_\mathrm{L}^\mathrm{in} +
{\mathcal{J}}_\mathrm{R}^\mathrm{out})\rho(t)\}$.  In the large-bias limit,
the superoperators $\mathcal{J}_\mathrm{L}^\mathrm{in} ,
\mathcal{J}_\mathrm{R}^\mathrm{out}$ act on the reduced density
operator according to
\begin{align}
\mathcal{J}_\mathrm{L}^\mathrm{in}\, \rho
={} & \Gamma_\mathrm{L} c_3^\dagger  \rho c_3 ,
\\
\mathcal{J}_\mathrm{R}^\mathrm{out}\, \rho
={} & \Gamma_\mathrm{R} c_4  \rho c_4^\dagger .
\end{align}
They describe incoherent tunnelling of an electron from the left lead
to dot $D_3$ and from dot $D_4$ to the right lead, respectively.
Notice that for large bias voltage, the meter properties no longer
depend on the absolute values of the onsite energies $\varepsilon_3$
and $\varepsilon_4$, but only on the energy bias $\varepsilon =
\varepsilon_4-\varepsilon_3$.
All numerical results presented below have been obtained by
integrating master equation \eqref{mast}.

\section{Measurement concept and visibility}
\label{sec:chargereadout}

The central idea of our readout scheme is that the current through the
open double dot depends on the position of the electron in the closed
quantum dot, i.e., on the state of the qubit.  The qubit in turn is
affected by the coupling to the leads via the meter which thus
effectively represents a macroscopic environment.  Therefore, the
qubit will experience decoherence, such that the qubit-meter setup
will evolve into a generally unique stationary mixed state in which
the current assumes a value independent of the initial condition.
This transition corresponds to a ``collapse'' of the wave function
during a finite time.  Moreover, it implies that the readout is
destructive, i.e., some information about the qubit state is
transfered to the leads.  The full quantum state of the lead, however,
is not accessible such that the measurement is quantum limited.
Nevertheless, measurement of macroscopic lead observables such as the
current is possible.
In the present case, the current exhibits transients that allow
one to draw conclusions on the qubit's \textit{initial state}
which we propose to read out.
In the following, we reveal the underlying relation between the
initial qubit state and the transient current.

As a natural and experimentally relevant initial condition, we assume
that in the beginning, the meter is not coupled to the qubit ($U=0$)
and stays in the corresponding stationary state which is a mixed
non-equilibrium state in which a current flows.\cite{on_initialcondition}
At this stage, the qubit is in a pure state which we parameterize on
the Bloch sphere as
\begin{equation}\label{Qubitstate}
|\psi (t=0)\rangle_{\rm q}
=\cos(\theta/2)\,|D_1\rangle
 +e^{i\phi}\, \sin(\theta/2)\,|D_2\rangle ,
\end{equation}
where the state $|D_{1,2}\rangle$ refers to an accordingly localized
electron.  The angle $\theta=0\ldots\pi$ determines the position
$\langle\sigma_z\rangle = \langle n_1\rangle-\langle n_2\rangle =
\cos\theta$ of the electron, while $\phi=0\ldots 2\pi$ denotes the
relative phase.
State preparation with such qubits has been demonstrated
experimentally in Refs.~\onlinecite{Hayashi2003a, Fujisawa2004a}.
At time $t=0$, the qubit-meter coupling $U$ is switched on, such that
the qubit influences the current and the readout process starts.

\begin{figure}[t!]
\centerline{\includegraphics{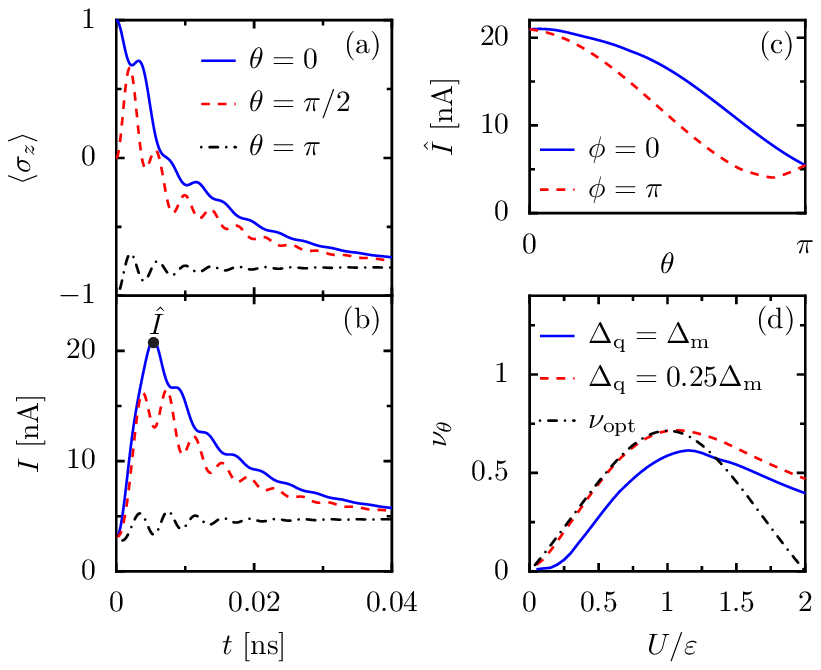}}
\caption{\label{I_t_Charge}
(Color online)
Charge readout for
$\Gamma_\mathrm{L}=\Gamma_\mathrm{R}=0.2\,\mathrm{meV}$,
$\Delta_\mathrm{q}=\Delta_\mathrm{m}=0.2\,\mathrm{meV}$, and
$U = \varepsilon = 1\,\mathrm{meV}$.
(a) Transient dynamics of the qubit population $\langle\sigma_z\rangle
= \langle n_1-n_2\rangle$ and (b) corresponding current for various
initial localizations.
(c) Height of the current peak as a function of the
initial occupation $\cos(\theta/2)$ for different relative phases $\phi$.
(d) Charge-readout visibility $\nu_\theta$.  The dashed-dotted
line marks the ``optimal'' visibility, for which backaction is ignored;
see Fig.~\ref{Ausleseidee}(b).}
\end{figure}%
Before addressing phase readout, we elucidate the underlying mechanism
for the more intuitive charge readout.\cite{Goan2001a, Gilad2006a,
Jiao2007a}  Figures~\ref{I_t_Charge}(a,b) show a typical time
evolution of the qubit population and the corresponding current.  If
the qubit is initially in state $|D_2\rangle$, i.e.\ for $\theta=\pi$,
it will essentially stay there and, thus, the open double dot remains
off-resonant and the current small.  This stems from the fact that for
a given $U$, the energy of the two double dots is smaller when the
qubit electron resides in one particular dot, which for the present
parameters is dot $D_2$.  This implies that the initial state
$|D_2\rangle$ is already close to the stationary state and not much
dynamics is going on.
When starting in state $|D_1\rangle$ ($\theta=0$), by contrast, the
capacitive coupling tunes the levels of dots $D_3$ and $D_4$ into
resonance and the current starts to increase until the systems evolves
into a stationary state with an again smaller current.  Thus, we
observe a current peak.  The solid line in Fig.~\ref{I_t_Charge}(c)
shows that the peak current $\hat I$, i.e.\ the maximal current, is
related to the population parameter $\theta$.  This clear dependence
is quite important for the readout scheme, because it implies  that
the measurement of $\hat I$ corresponds to the determination of the
expectation value $\langle\sigma_z\rangle = \cos\theta$ for the
initial qubit state.

Readout with good resolution requires that the peak height
depends strongly on the initial population.  A figure of merit for this is
the scaled difference between the maximum and the minimum of  $\hat I$
upon variation in $\theta$, i.e., the charge-readout visibility
defined as
\begin{equation}
\label{vis-charge}
\nu_\theta = \frac{\max_\theta \hat I - \min_\theta\hat I}
                  {\max_\theta\hat I + \min_\theta\hat I} \,.
\end{equation}
Figure~\ref{I_t_Charge}(d) shows this quantity for two different qubit
splittings.  The visibility is best ($\nu_\theta\approx 0.75$) when the
interaction energy $U$ matches the bias $\varepsilon$ of the meter and
the qubit dynamics is slower than electron tunnelling through the
meter, i.e., when $\Delta_\mathrm{q} < \Delta_\mathrm{m},
\Gamma_\mathrm{L,R}$.  For $U\lesssim\varepsilon$, the visibility even
reaches the optimal value which is achieved when backaction of the
meter to the qubit is ignored.  This complies with the picture drawn
by investigating the peak-to-background ratio.\cite{Gilad2006a,
Jiao2007a}

Most interestingly, Fig.~\ref{I_t_Charge}(c) indicates that the peak
current $\hat I$ for a given initial population also depends on the initial
relative phase $\phi$.  This already hints on the feasibility of phase
readout, despite the fact that the difference in $\hat I$ is not very
pronounced for the parameters used here.

\section{Phase-readout}
\label{sec:phasereadout}

We have already seen in the last section that phase readout is
possible in principle.  This raises two intriguing questions.
First, we would like to qualitatively understand why phase readout
works, despite the fact that the capacitive coupling is sensitive to
the location of the qubit electron only.  The second question is of
quantitative nature: can one achieve a visibility comparable to the
one obtained for charge readout?

The relative phase is only meaningful when both qubit states are
populated, and a noticeable influence on the transient current
requires even significant population of both states.  Therefore, we
restrict ourselves to equal initial population, i.e., to $\theta =
\pi/2$.
In order to gain a physical picture of the phase readout, let us focus
on the two phases $\phi=\pi$ and $\phi=0$, i.e., on the states
$(|D_1\rangle-|D_2\rangle)/\sqrt{2}$ and
$(|D_1\rangle+|D_2\rangle)/\sqrt{2}$, which are
the eigenstates of the qubit Hamiltonian~\eqref{Hq} in the one-electron
subspace.  Since the qubit couples via the meter to a
macroscopic environment, it will evolve into an asymptotic state.
Thus, for the ground state ($\phi=\pi$), the qubit will absorb energy
from the meter, while for the excited state ($\phi=0$), the qubit
emits energy.  Both processes leave their fingerprints in the
transients of the current and, thus, allow one to discern different
initial phases.

\begin{figure}[t!]
\centerline{\includegraphics{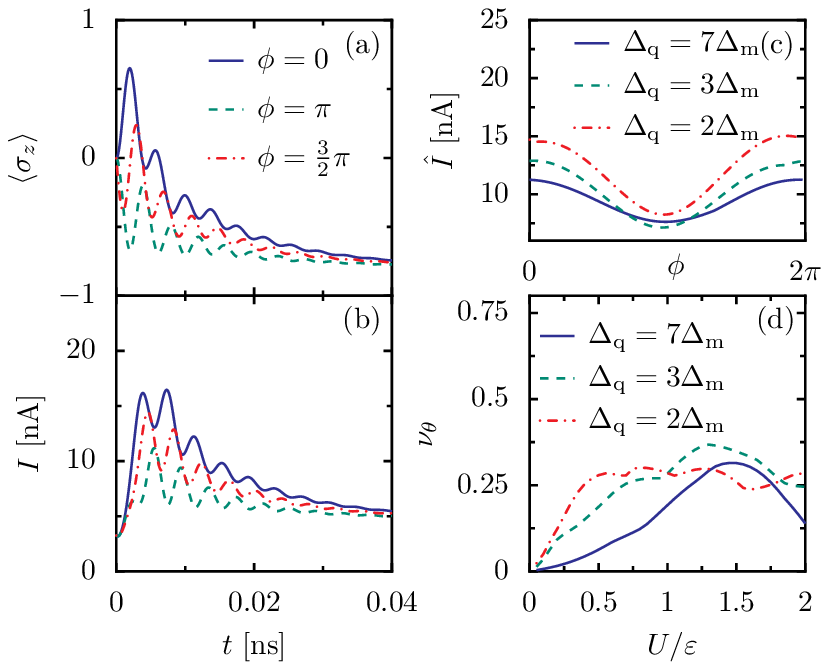}}
\caption{\label{I_t_Phase}
(Color online)
Phase readout of a qubit with energy splitting $\Delta_\mathrm{q} =
0.2\,\mathrm{meV}$ coupled to a meter with the same parameters as
in Fig.~\ref{I_t_Charge}.  The qubit states are at initial
time equally populated, i.e., $\theta=\pi/2$.
(a) Transient qubit dynamics and (b) current for various initial
phases $\phi$ and qubit-meter interaction $U=\varepsilon = 1\,
\mathrm{meV}$ and $\Delta_\mathrm{q}=0.2\,\mathrm{meV}$.
(c) Height of the current peak as a function of the initial phase.
(d) Readout visibility as a function of the qubit-meter interaction
strength~$U$.
}
\end{figure}

A typical transient current is shown together with the corresponding
qubit dynamics in Figs.~\ref{I_t_Phase}(a,b).  The current peak
resembles the one analyzed in the context of charge readout.  Here
however, the peak never vanishes completely, as it was the case above
for the initial state $|D_2\rangle$.  Thus, at first glance, phase
readout seems to possess a much lower resolution than charge readout.
For a quantitative analysis we consider the \textit{phase-readout
visibility} $\nu_\phi$
which we define according to Eq.~(\ref{vis-charge}) but with the
population parameter~$\theta$ replaced by the relative phase~$\phi$.

Figure~\ref{I_t_Phase}(c) reveals the clear $\phi$-dependence of the
current peak height $\hat I$,  which forms the basis of our
phase-readout scheme.  The maximal value and the minimal value of
$\hat I$ are assumed for phases very close to $\phi=0$ and $\phi=\pi$,
respectively.  Therefore the experimental distinction between these
two phases can be achieved with the full visibility $\nu_\phi$ shown
in Fig.~\ref{I_t_Phase}(d).  As compared to charge readout, the
visibility exhibits a less regular structure.  For the
interaction strength $U\approx \varepsilon/2$, it reaches a value
$\nu_\phi \approx 0.25$ and remains of that order when $U$ is
increased.  In contrast to charge readout, we find a tendency toward
higher visibility for larger qubit splitting.  Nevertheless, for these
parameters, $\nu_\phi$ still stays clearly below the charge readout
visibility $\nu_\theta$.  Therefore it is essential to optimize the
setup.

Three routes toward an optimized phase readout come to mind.
First, as already noticed above, the qubit splitting
$\Delta_\mathrm{q}$ should be larger than the tunnel matrix element
$\Delta_\mathrm{m}$ of the meter.  Irrespective of any experimental
constraints, increasing $\Delta_\mathrm{q}$ is only of limited use,
because beyond a certain limit, the qubit oscillations then become so fast
that the meter is no longer able to follow.  Consequently, the meter
no longer contains information on the qubit and, thus, the readout
quality will decrease.  In our case, we find that $\Delta_\mathrm{q} =
3\Delta_\mathrm{m}$ is a good choice, while for larger qubit
splittings, the visibility indeed decreases; see Fig.~\ref{I_t_Phase}(d).

A second way for improving the visibility is to tune the
meter into a regime of higher sensitivity which is mainly
determined by the resonance curve shown in Fig.~\ref{Ausleseidee}(b).
When the dot-lead tunnel rates $\Gamma_\mathrm{L,R}$ become smaller,
the current maximum $I_1$ increases, while the current for an energy
bias $\varepsilon=U$, which is $I_2$, decreases.  Consequently, the
achievable visibility $\nu_\mathrm{opt}$ becomes larger.  The reduced
dot-lead tunnelling, however, leads to a smaller current, such that
the current measurement eventually will be difficult.

Alternatively, one can reduce the current $I_2$ by using a setup with
a larger bias $\varepsilon$ and an accordingly larger interaction
energy $U$.
The Coulomb interaction $U$, however, is determined by the distance
between the dots $D_1$ and $D_3$ and, thus, is limited by the size of
the top gates that define the quantum dots.  Nevertheless, it is
possible to enhance the qubit-meter coupling by choosing a setup
in which also the electrons in dots $D_2$ and $D_4$ repel each other.
Then the qubit-meter Hamiltonian (\ref{WireQubit}) has to be extended
by the term $U' n_2 n_4$.  If now an electron tunnels from $D_2$
to $D_1$, the left meter level is raised by $U$, while the right meter
level is no longer raised by $U'$, i.e., it is effectively
lowered by $U'$.  This implies that the relevant effective interaction
strength is $U_\mathrm{eff} = U+U'$.  Notice that the resonance condition
for the meter bias nevertheless reads $\varepsilon = U$.  This
additional qubit-meter coupling represents our third way of optimization.
We explore its benefits by comparing two situations with the same
effective coupling strength: symmetric coupling $U=U' =
U_\mathrm{eff}/2$ and coupling only between dots $D_1$ and $D_3$,
which means $U=U_\mathrm{eff}$ while $U'=0$.

\begin{figure}[t!]
\centerline{\includegraphics{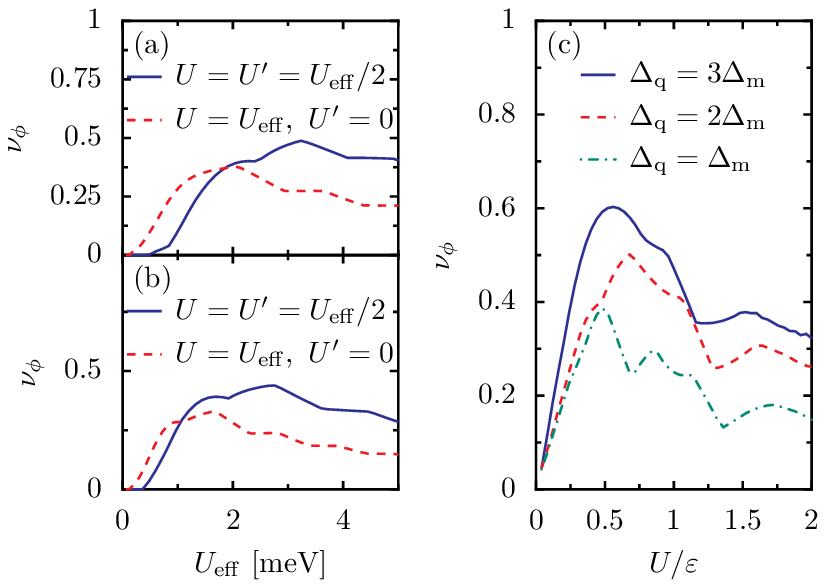}}
\caption{\label{VisPhase}
(Color online)
Phase-readout visibility for a meter with $\Gamma_{\rm L} =\Gamma_{\rm
R} =0.2\ {\rm meV}$ and tunnel matrix element $\Delta_\mathrm{m}
=0.2\ {\rm meV}$.
Comparison of symmetric coupling ($U=\varepsilon$) and coupling
between the left dots ($U'=0$) for qubit energy splitting (a)
$\Delta_\mathrm{q} = 3\Delta_\mathrm{m}$ and (b) $\Delta_\mathrm{q} =
2\Delta_\mathrm{m}$.
(c) Visibility as a function of the interaction strength $U$ for
symmetric coupling $U'=U$, bias $\varepsilon=1.5\,{\rm meV}$ and
various qubit splittings.
}
\end{figure}%
Figures~\ref{VisPhase}(a,b) show the resulting visibilities as a
function of the meter bias.  We find that in the relevant regime with
large $\nu_\phi$, the symmetric coupling is superior to the asymmetric
coupling.  The difference is up to roughly 30\%.
Finally, in order to explore the limits of the symmetrically coupled
setup, we plot in Fig.~\ref{VisPhase}(c) the visibility as a function
of the interaction strength.  It turns out that the result is best in
the vicinity of $U=\varepsilon/2$.
The observed maximum is rather broad, which means that our readout scheme
is quite tolerant against moderate parameter variations.  Summarizing our
optimization procedure, we find that the phase-readout visibility can
be up to $\nu_\phi = 0.6$, which is smaller but still of the same
order as what we found for charge readout ($\nu_\theta=0.75$).  This
clearly  demonstrates the feasibility of phase readout.

For the numbers used in our study, the current peaks are on
the order of $10\,\mathrm{nA}$ with a duration of up to
$10\,\mathrm{ns}$.  Thus a peak typically consists of ten electrons.
Assuming that the transported electrons are uncorrelated, they obey
Poisson statistics and, thus, the standard deviation of the number of
transported electrons is $\pm3$.  This deviation should represent the
relevant fluctuation of the signal.
Owing to these numbers, the experimental realization of our phase
readout (and of related charge-readout schemes \cite{Wiseman2001a,
Gilad2006a, Jiao2007a} as well) represents a demanding task.
Nevertheless, it should be feasible, e.g., by repeating preparation
and readout many times, such that the current peaks turn into an
enhanced average current.\cite{Hayashi2003a, Fujisawa2004a}
Alternatively, one could think of combining the charge meter with
setups for monitoring the transport of individual
electrons.\cite{Gustavsson2006a, Fricke2007a}

At the readout stage of a quantum algorithm, one already knows that
the qubit resides in either of two specific orthogonal
states.\cite{Nielsen2000a}  If these states are the charge states
$|D_1\rangle$ and $|D_2\rangle$, the results of
Sec.~\ref{sec:chargereadout} allow one to distinguish between them.
The results of the present section, by contrast, are of use when the
two possible final states of a gate operation are given by
Eq.~\eqref{Qubitstate} for $\theta = \pi/2$ and for example $\phi =
0,\pi$.

\section{Conclusions}

We have investigated the transient current through an open double
quantum dot with a capacitive coupling to a charge qubit.  In
particular, we focused on the impact of the initial qubit state on
the current peak that emerges after the qubit is coupled to the open
double quantum dot.  Such qubit-meter setups have recently been
proposed for monitoring the location of an electron in a closed double
dot, i.e., for charge readout with single or double dot
meters.  Our results demonstrate that
such a charge meter is useful for phase readout as well, despite the
fact that the qubit phase possesses only indirect influence on the
measured current.  For an unbiased qubit, the relative phase
determines whether the qubit is initially in its ground state or in
its excited state.  Thus, when coupled to a macroscopic device such as
the meter, the qubit will absorb or emit energy, depending on its
initial phase.  This difference is visible in the height of a
transient current peak, whose measurement thus corresponds to phase
readout.

After having realized that phase readout is possible, in principle, we
have investigated whether the measured signals are sufficiently
pronounced, such that they allow one
to reliably discern between different initial phases.  Our
results for the phase readout visibility, defined as the scaled
difference of the current peaks, reveal that phase readout is only
slightly more demanding than the previously proposed charge readout
with single or double quantum dot meters.
In conclusion, we believe that the experimental implementation of our
phase-readout scheme opens a promising way for the observation of
coherent tunnelling dynamics in double quantum dots.

\begin{acknowledgments}
This work has been supported by the DFG through SFB 631 and by the
German Excellence Initiative via ``Nanosystems Initiative Munich
(NIM)''.  S.K. acknowledges support by the Spanish Ministerio de
Ciencia e Innovaci\'on through the Ram\'on y Cajal program.
\end{acknowledgments}




\begin{thebibliography}{10}
\bibitem{on_initialcondition}
We also performed calculations with other initial conditions, such as an empty
  meter double quantum dot. Irrespective of their experimental feasibility, it
  is interesting to note that the results are similar to those presented below
  and lead to the same conclusions.
\end{thebibliography}

\end{document}